\documentclass[12pt]{article}

\usepackage{amssymb}
\usepackage{amsmath}
\usepackage{epsfig}
\usepackage{psfrag}

\def\be{\begin{equation}}
\def\ee{\end{equation}}
\def\bea{\begin{eqnarray}}
\def\eea{\end{eqnarray}}

\begin{document}
\begin{titlepage}
\thispagestyle{empty}
\hskip 1 cm
\vskip 0.5cm

\vspace{25pt}
\begin{center}
    { \LARGE{\bf Islands in the Landscape}}
    \vspace{33pt}

  {\large  {\bf   T. Clifton\thanks{T.Clifton@cantab.net}, Andrei Linde\thanks{alinde@stanford.edu}
  and Navin Sivanandam\thanks{navins@stanford.edu}}}

    \vspace{15pt}

 {Department of Physics,
    Stanford University, Stanford, CA 94305}

  \end{center}

   \vspace{20pt}

\begin{abstract}

The string theory landscape consists of many metastable de Sitter
vacua, populated by eternal inflation.  Tunneling between these
vacua gives rise to a dynamical system, which asymptotically settles
down to an equilibrium state.  We investigate the effects of sinks
to anti-de Sitter space, and show how their existence can change
probabilities in the landscape.  Sinks can disturb the thermal
occupation numbers that would otherwise exist in the landscape and
may cause regions that were previously in thermal contact to be
divided into separate, thermally isolated islands.

\end{abstract}

\vspace{10pt}
\end{titlepage}

\tableofcontents

\newpage

\section{Introduction}

One of the most intriguing features of string theory is its
prediction of a multitude of vacuum states
\cite{Lerche:1986cx,Bousso:2000xa,Douglas,Denef:2007pq,Danielsson:2006xw}
 Stabilizing these states
\cite{Kachru:2003aw} and coupling the resulting embarrassment of
riches with a population mechanism \cite{book,LLM} provided by
eternal inflation \cite{Vilenkin:1983xq,Eternal} gives rise to the
string theory landscape \cite{Susskind:2003kw}. Physics in the
landscape can be both rich and perplexing.  We aim to elucidate some
relevant ideas for understanding this physics, with the hope of
improving our comprehension of the multiverse.

Inflationary expansion divides the universe into many exponentially
large domains, each corresponding to different metastable vacuum
states.  In this picture tunneling between different vacuum states
causes bubbles of new vacuum to be continually nucleated. Those with
positive vacuum energy are initially static, but soon accelerate in
their expansion until the velocity of their walls asymptotically
approaches that of light.  If the conditions are right inside these
bubbles then a stage of  slow-roll inflation  will occur and the resulting observers will
see themselves in an infinitely extended, open Friedmann universe.
Percolation of successive bubbles inside of each other give us a
universe that is eternally inflating and constantly producing new
inflationary universes, where structure can form, and life can evolve.

The first step in a complete understanding of this scenario must be
to find out which vacua are possible in string theory, and to
describe their typical properties \cite{Douglas}.  Once these vacua
have been identified we will then need to study cosmological
evolution during eternal inflation in order to determine the global structure of
the   universe \cite{LLM}.  During inflation the number of
horizon-sized dS regions of space-time is continually, and
exponentially, increasing.  These dS regions then `populate' the
many possible vacua of string theory, realizing the great variety of
the theory in a diverse and eternal universe.

The resulting picture is incredibly complex. Ultimately, our goal is
to explain the properties of our part of the multiverse, and to
predict the results of future observations. To achieve this goal one
needs to calculate the probabilities of various outcomes in an
eternally inflating multiverse. This is a thorny problem, and a
subject of much contention. Studying the global structure of an
eternally inflating spacetime leads to comparisons of infinite
volumes, and hence a consequent dependence on cutoff procedures
\cite{LLM,Bellido,Mediocr,Vilenkin:2006qf,LMprob,Tegmark:2004qd,Aguirre:2006ak}.
Escaping cutoff problems is not impossible if one considers
individual observers and concentrates on their individual histories,
ignoring the rest of the universe
\cite{Starobinsky:1986fx,Goncharov:1987ir,Garriga:2005av,Bousso:2006ev,Vachurin:2006};
here, however, we must face the problems of initial conditions and
are, perhaps, led to worry about Euclidean quantum gravity and the
wave-function of the universe
\cite{Hawking:2006ur,Hartle:1983ai,Linde:1983mx}. More importantly,
this description tends to miss some of the important features of
eternal inflation.

In this paper we will leave anthropic considerations aside and
concentrate on other properties of the string theory landscape. We
will focus on the existence, or otherwise, of thermal equilibrium
between populations of dS vacua. As we will see, under certain
conditions, a system of dS vacua  described in comoving coordinates
settles down to a state in which the populations of these vacua are
in thermal equilibrium with one another. More precisely, the ratios
of comoving volume occupied by one vacuum or another will depend on
the exponential of the entropy difference between them.  An
important limitation of this simple picture is that it is valid only
in comoving coordinates, which do not reward different rates of
cosmological expansion in different parts of the universe.
Nevertheless, the picture of many dS universes in a state of thermal
equilibrium is very simple and intuitively appealing, and therefore
it can be very useful for understanding various features of the
string theory landscape.

On the other hand, this simple picture may be invalid when the landscape has sinks
(terminal vacua which can be tunneled to, but not from). In particular, in
\cite{Ceresole:2006iq,Linde:2006nw} it was shown that for a simple
system consisting of 2 dS vacua and one AdS sink, naive expectations
of thermal equilibrium are incorrect if the decay rate to the sink
is sufficiently fast. Since one expects sinks to be common in the
landscape \cite{Ceresole:2006iq}, it may be the case that the
disruption of thermal equilibrium between metastable dS vacua is a
generic feature, and the string theory landscape may consist of many
thermally isolated `islands.' The goal of this work is to elucidate
this possibility and to investigate in more detail the situations in
which the usual thermally equilibrium populations are disturbed.  We will
find explicit solutions for a variety of simple configurations that
may occur in the landscape, and use these results to form a picture
of how the vacua of a more realistic landscape may be populated.

It will be found that the presence of sinks in the landscape can
significantly alter the dynamics of the inflating multiverse.  One
of the most dramatic and unexpected examples of this is that when a
number of high energy vacua decay to a single lower energy vacuum,
which can decay to a sink, the probability fluxes soon become
dominated by the slowest decaying, most stable vacuum. In the
limiting case of this vacuum being completely stable it makes no
contribution at all to probability fluxes, as these are the results
of tunneling events between metastable vacua (see below).  However,
if the smallest chance of tunneling out of this vacuum is allowed, we
unexpectedly find that this tiny current comes to be the dominant
source of the probability flux. This slowest decaying vacuum may then
remain out of thermal contact with other vacua, whilst all faster
decaying vacua eventually approach thermal equilibrium with each
other.  This `tortoise and the hare' scenario shows explicitly the
non-trivial effect of sinks on the dynamics of inflation in the string
theory landscape: They may lead to the existence of thermally isolated,
slowly decaying vacua while all other, more rapidly decaying vacua
are left in thermal equilibrium.

We will also investigate the possibility that some of the  AdS or
Minkowski sinks could potentially act as impassable barriers between
systems of dS vacua, thus carving the landscape into totally
disconnected `islands'.    Such a
situation would result in different regions of the multiverse being
completely isolated from one another, whilst maintaining
thermal equilibrium internally.   We argue that the large number of
vacua and dimensions in the landscape, coupled with the `vacuum
dynamics' we find to exist in the presence of sinks, makes the
existence of such isolated regions improbable, though perhaps not impossible.

In section \ref{dS} we will review the basic mechanisms of tunneling
between vacua. Following this, in section \ref{tunnsink} we will
outline the results of \cite{Ceresole:2006iq,Linde:2006nw}, for a
simple landscape of two dS vacua and one AdS sink.
Section \ref{newsec} discusses the possibility
of sinks dividing the landscape into
disconnected islands.
In section \ref{island} we shall summarize the results of our
investigations into more complicated toy landscapes, highlighting
some of the counter-intuitive features that emerge. In section
\ref{disc} we summarize our results.  Mathematical details can be
found in the appendix.

\section{Tunneling in the Landscape}\label{dS}

There are two related mechanisms for making transitions between
vacua: one due to tunneling \cite{Coleman:1980aw} and another due to
stochastic diffusion processes
\cite{Starobinsky:1986fx,Linde:1991sk}. A somewhat more detailed
discussion of these mechanisms, and the issues associated with them,
can be found in \cite{Linde:2006nw}. We summarize the salient points
below.

Tunneling between vacua produces bubbles of new vacuum, that look
like infinite  open Friedmann universes to observers
inside. If the tunneling goes to dS space, then the bubble expands
exponentially with the velocity of its walls approaching that of
light.  (In comoving coordinates these bubbles approach some maximal
value and freeze. This maximal value depends on the time when the
bubble is formed, and is exponentially smaller for bubbles formed
later on \cite{Guth:1982pn}.) If the tunneling goes to a state with
a negative vacuum energy $V$, the infinite universe inside it
collapses within a time of the order $|V|^{-1/2}$, in Planck units.
These negative energy, AdS vacua then play the role of sinks for
probability currents in the landscape.

\psfrag{phi1}{$\phi_1$} \psfrag{phi2}{$\phi_2$}
\psfrag{phitop}{$\phi_{top}$}
\begin{figure}[ht]
\center \epsfig{file=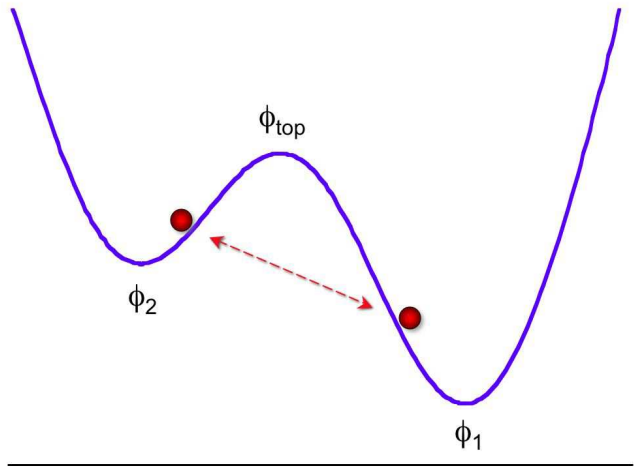,height=6.5cm}
\caption{{\protect {\textit{Coleman-De Luccia
  tunneling may go in both directions.}}}}
\label{2min}
\end{figure}

Let us consider two dS vacua, $\mathrm{dS}_{i}$, with vacuum energy
density $V_{i} = V(\phi_{i})$, Fig. \ref{2min}. Without taking
gravity into account, the tunneling may go only from the upper
minimum to the lower minimum, but in the presence of gravity
tunneling may occur in both directions, which is emphasized in Fig.
\ref{2min}. According to Coleman and De Luccia
\cite{Coleman:1980aw}, the tunneling probability from
$\mathrm{dS}_{1}$ to $\mathrm{dS}_{2}$ is given by
\begin{equation}
\label{prob} \Gamma_{12} = e^{-B} = e^{-S(\phi)+S_1},
\end{equation}
where $S(\phi)$ is the Euclidean action for the tunneling
trajectory, and $S_1=S(\phi_1) $ is the Euclidean action for the
initial configuration $\phi = \phi_1$,
\begin{equation}
\label{action2} S_1 = - {\frac{24\pi^2}{ V_1}} <0\ .
\end{equation}
This action has a simple sign-reversal relation to the entropy of de
Sitter space, ${\bf S_1}$:
\begin{equation}
\label{action2a} {\bf S_1} = - S_1 =+ {24\pi^2\over V_1}\ .
\end{equation}
Therefore the decay time of the metastable dS vacuum  $t_{\rm decay}
\sim \Gamma^{-1}_{12}$ can be represented in the following way:
\begin{equation}
\label{decaytime} t_{\rm decay} = e^{S(\phi)+\bf S_1} = t_r \ e^{S(\phi)}\ .
\end{equation}
Here $t_{r} \sim e^{\bf S_{1}}$ is the so-called recurrence time for
the vacuum $\mathrm{dS}_{1}$.

Whereas the theory of tunneling developed in \cite{Coleman:1980aw} was
quite general, all examples of tunneling studied there described the
thin-wall approximation, where the tunneling occurs from one  minimum
of the potential and proceeds directly to another minimum. In the cases where the thin-wall approximation is not valid, the
tunneling occurs not from the minimum but from the wall, which makes
interpretation of this process in terms of the decay of the initial
vacuum less trivial.

The situation becomes especially confusing when the potential is very flat on the way from
one minimum to another, $V'' < V$, in Planck units. In this case the
Coleman-De Luccia (CDL) instantons describing decay of a dS space do not
exist \cite{Hawking:1981fz}; they become replaced by Hawking-Moss (HM)
instantons.  According to Hawking and Moss
\cite{Hawking:1981fz}, the probability of tunneling from the minimum
1 to the minimum 2 is then given by
\begin{equation}
\label{HM} \Gamma_{12} = e^{-S_{\rm top}+S_1}=
\exp\left(-{24\pi^2\over V(\phi_{1})}+{24\pi^2\over V(\phi_{\rm
    top})}\right) \ .
\end{equation}
The HM instanton is described by the Euclidean version of dS space corresponding
to the top of the potential barrier, $\phi = \phi_{\rm top}$.

Unlike the thin-wall CDL solution, the HM solution does not
interpolate between the two different minima of $V(\phi)$, and
therefore debates on the validity of the HM result continue even now
\cite{Weinberg:2006pc}. One may wonder why we should consider
such instantons instead of considering the  instantons corresponding
to the dS space in the next minimum; the resulting tunneling action
would be much smaller. Moreover, one may consider a string theory
landscape with many minima and maxima separated by a sequence of
barriers. Then one could wonder whether the HM tunneling suppression
applies only to the tunneling between the nearby vacua, or if it can
describe direct tunneling to distant minima, ignoring all
intermediate barrier except the last one \cite{book,Weinberg:2006pc}.
One of the best attempts to clarify this situation was made by Gen and
Sasaki \cite{Gen:1999gi}, who described the tunneling using
Hamiltonian methods in quantum cosmology, which avoided many
ambiguities of the Euclidean approach. But even their investigation
does not allow us to answer the last of these questions.

A proper interpretation of the Hawking--Moss tunneling was achieved
only after the development of the stochastic approach to inflation
\cite{LLM,Starobinsky:1986fx,Linde:1991sk,Goncharov:1986ua}. One may
consider quantum fluctuations of a light scalar field $\phi$ with
$m^2 = V'' \ll H^2 = V/3$.  During each time interval $\delta t =
H^{-1}$ this scalar field experiences quantum jumps with the
wavelength $\sim H^{-1}$, and with a typical amplitude $\delta\phi =
H/2\pi$. As a result, quantum fluctuations lead to a local change
in amplitude of the field $\phi$, which looks homogeneous on the
horizon scale $H^{-1}$. From the point of view of a local observer,
this process looks like a Brownian motion of the homogeneous scalar
field. If the potential has a dS minimum at $\phi_1 \gg H^{2}/m$,
then eventually the probability distribution to find the field with
the value $\phi$ at a given point becomes (almost) time-independent,
\begin{equation}
\label{E38a} P(\phi) \sim \exp\left(-{24\pi^2\over V(\phi_{1})}+{24\pi^2\over V(\phi)}\right) \ .
\end{equation}

The distribution $P(\phi)$ gives the probability to find the
field $\phi$ at a given point, and has a simple interpretation as the
fraction of {\it comoving} volume of the universe in each of the dS vacua, or, equivalently, a fraction of time the field spends in a vicinity of its value $\phi$ along the Brownian trajectory.
Up to a sub-exponential factor, this  distribution shows the density of points with a given value of the field $\phi$ along its Brownian trajectory. This implies that, up to a sub-exponential factor,  the typical time
required for the field,  at any given
point in comoving coordinates, to move from its equilibrium value $\phi_{1}$ and climb to the top of the barrier is proportional to
$P^{-1}(\phi_{\rm
    top}) \sim \exp\left({24\pi^2\over V(\phi_{1})}-{24\pi^2\over V(\phi_{\rm
    top})}\right)$ \cite{book}.
 Once the scalar field climbs to the top of the
barrier, it can fall from it to the next minimum, which completes
the process of ``tunneling'' in this regime. That is why the
probability to gradually climb to the local maximum of the potential
at $\phi = \phi_{\rm top}$ and then fall to another dS minimum is
given by the Hawking-Moss expression (\ref{HM})
\cite{Starobinsky:1986fx,Goncharov:1986ua,book,Linde:1991sk}. It is
also why tunneling to distant minima separated by many barriers is
accomplished by a sequence of transitions from one minimum to another
nearby minimum, rather than by one big jump. This last statement does
not follow from the Hawking-Moss derivation of their result, but is
apparent from the stochastic approach to inflation.

A necessary condition for the derivation of Eq. (\ref{E38a}) using the  stochastic approach to inflation in
\cite{LLM,Starobinsky:1986fx,Linde:1991sk,Goncharov:1986ua} is the
requirement that $m^2 = V'' \ll H^2 = V/3$.  This requirement is
satisfied during slow-roll inflation, but it is violated for all known
scalar fields at the present (post-inflationary) stage of the evolution of the universe. Thus the
situation with the interpretation of the Coleman-De Luccia tunneling
for $ V'' \geq   V/3$ is somewhat unsatisfactory. However, since the
validity of the Coleman-De Luccia approach was confirmed at least in
some limiting cases (in the absence of gravity, and in the slow-roll
regime discussed above), in this paper we will follow the standard
lore, assume that this approach is correct, and study its consequences.

Following \cite{Lee:1987qc} (see also
\cite{Susskind:2003kw,Garriga:1997ef,Dyson:2002pf}), we will look
for the probability distribution $P_{i}$ to find a given {\it point}
in a state with vacuum energy $V_{i}$, and will try to generalize
the results for the probability distribution obtained above by  the
stochastic approach to inflation.  The main idea is to consider CDL
tunneling between two dS vacua, with vacuum energies $V_{1}$ and
$V_{2}$, such that $V_{1} < V_{2}$, and to study the possibility of
tunneling in both directions, from $V_{1}$ to $V_{2}$, or vice
versa.

The action on the tunneling trajectory, $S(\phi)$, does not depend
on the direction in which the tunneling occurs, but the tunneling
probability does depend on it. It is given by $e^{-S(\phi)+S_1}$ on
the way up, and by $e^{-S(\phi)+{ S_2}}$ on the way down
\cite{Lee:1987qc} (for a recent discussion of related subjects see also \cite{Aguirre:2005nt}). Let us  assume that the universe is in a
stationary state, such that the comoving volume of the parts of the
universe going upwards is balanced by the comoving volume of the
parts going down. This can be expressed by the detailed balance
equation
\begin{equation}
\label{balance} P_{1}\, e^{-S(\phi)+S_1} = P_{2}\, e^{-S(\phi)+S_2} \ ,
\end{equation}
which yields (compare with Eq.~(\ref{HM}))
\begin{equation}
\label{weinb} {P_{2}\over P_{1}} = e^{-S_{2}+S_1} = \exp\left(-{24\pi^2\over V_1}+{24\pi^2\over V_2}\right) \ ,
\end{equation}
independently of the tunneling action $S(\phi)$.

This probability distribution also has a nice thermodynamic
interpretation in terms of dS entropy  ${\bf S}$
\cite{Linde:1998gs}:
\begin{equation}
\label{weinb2} {P_2\over P_{1}} = e^{{\bf S_{2}}-{\bf S_{1}}} = e^{{\bf \Delta S}}\ .
\end{equation}
Here, as before, ${\bf S_{i}} = - S_{i}$. This result and its
thermodynamic interpretation have played a substantial role in the
discussion of the string theory landscape \cite{Susskind:2003kw}.

Following \cite{Ceresole:2006iq,Linde:2006nw} we shall now show this
notion of thermal equilibrium is a fragile one, and can be destroyed
by a simple extension of our model to include sinks.

\section{A Toy Landscape: 2 dS and 1 AdS}\label{tunnsink}

Stationarity of the probability distribution (\ref{weinb2}) was
achieved because the lowest dS state did not have anywhere further
to fall.  Meanwhile, in string theory all dS states are metastable,
so it is always possible for a  dS vacuum to decay
\cite{Kachru:2003aw}.  Further, it is important that if it decays by
the production of bubbles of 10D Minkowski space, or by production
of bubbles containing a collapsing open universe with a negative
cosmological constant, then the standard mechanism of returning back to
the original dS state  no longer operates. Therefore Minkowski vacua, as well as AdS vacua, work like
sinks for the flow of probability in the landscape. Because of the
existence of these sinks (also known as terminal vacua), the
fraction of the comoving volume in the dS vacua will decrease in
time.

Although decays from one SUSY AdS vacua to another are forbidden,
uplifting \cite{Kachru:2003aw} breaks supersymmetry. Uplifted dS
vacua can then decay to AdS by the formation of bubbles of
collapsing universes. According to \cite{Ceresole:2006iq},  the
typical decay rate for this process can be estimated as  $\Gamma
\sim \exp\left({C
  M_{p}^{2}/m^{2}_{3/2}}\right)$. For a gravitino mass, $m_{3/2}$, in the 1 TeV
range one finds suppression in the range of $\Gamma \sim
10^{-{10^{34}}}$  \cite{Ceresole:2006iq}, which is much greater than
the expected rate of decay to Minkowski vacua, or to higher dS
vacua, which (in vacua like ours) is typically suppressed by
factors of order $10^{-10^{120}}$. Other possible decay channels for
the uplifted dS space were discussed in
\cite{Frey:2003dm,Green:2006nv,Danielsson:2006jg}.

\psfrag{1}{$1$} \psfrag{2}{$2$} \psfrag{S}{$S$}
\begin{figure}[ht]
\center \epsfig{file=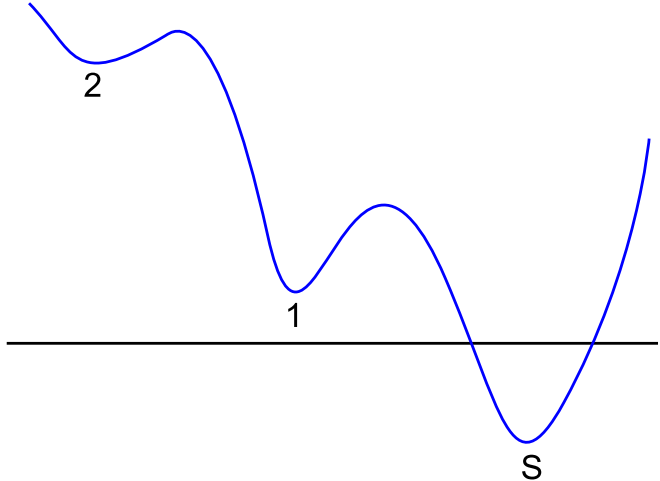,height=6.5cm}
\caption{{\protect {\textit{A potential with two dS minima and a sink.}}}}
\label{2ds1ads}
\end{figure}

Let us consider a simple model describing two dS minima and one AdS
minimum (denoted by 1, 2, and S in Fig. \ref{2ds1ads}). Here, as in
the rest of this paper, we work in comoving co-ordinates. To get a
visual understanding of the process of bubble formation in comoving
coordinates, one may paint black all of the parts corresponding to
one of the two dS states, and paint white the parts in the other dS
state. Then, in the absence of  sinks in the landscape, the
multiverse will become populated by white and black bubbles of all
possible sizes. Asymptotically, it will approach a stationary regime
-- on average becoming gray, with the level of gray becoming asymptotically
constant.  Suppose now that some parts of the universe may tunnel to a state
with a negative cosmological constant. These parts will collapse, so
they will not return to the initial dS vacua. If we paint such parts
red, then the universe, instead of reaching a constant shade of
gray, eventually will look completely red.

To describe this process, instead of the detailed balance equation
(\ref{balance}) one should use the  ``vacuum dynamics'' equations
\cite{Garriga:2005av,Ceresole:2006iq}:
\begin{equation}
\label{v01} \dot P_{1} = - J_{1s} -J_{12} + J_{21} \ ,
\end{equation}
\begin{equation}
\label{v02} \dot P_{2} = - J_{2s} -J_{21}+J_{12} \ .
\end{equation}
Here $J_{ij} = P_{j}\, \Gamma_{ji}$, where $ \Gamma_{ji}$ is the decay
rate of the vacuum $j$ to bubbles of vacuum $i$. In particular, $ J_{1s} = P_{1}\,e^{-C_{1}}$ is the
probability current from the lower dS vacuum to the sink,  i.e. to a
collapsing universe, or to a Minkowski vacuum, $ J_{2s} = P_{2}\,
e^{-C_{2}}$  is the probability current from the upper dS vacuum to
the sink, $ J_{12} =P_{1}\, e^{-{\bf S_{1}}+|S(\phi)|}$ is the
probability current from the lower dS vacuum to the upper dS vacuum,
and $ J_{21} = P_{2}\, e^{-{\bf S_{2}}+|S(\phi)|}$ is the probability
current from the upper dS vacuum to the lower dS vacuum. Combining
this all together, gives us the following set of equations for the
probability distributions:
\begin{equation}
\label{v1} \dot P_{1} =  -P_{1}\, (\Gamma_{1s} +\Gamma_{12}) +  P_{2}\,\Gamma_{21} \ ,
\end{equation}
\begin{equation}
\label{v2} \dot P_{2} =  -P_{2}\, (\Gamma_{2s} +\Gamma_{21}) +  P_{1}\,\Gamma_{12} \ .
\end{equation}
We ignore here possible sub-exponential corrections, which appear,
e.g., due to the difference in the initial size of the bubbles etc.

Because of the decay to the sink, $P_{1}$ and $P_{2}$ gradually become exponentially small. But this does not mean that the whole universe goes to the sink: The physical volume of white and black parts of the universe continues growing exponentially, in the regime of eternal inflation. One of the ways to account for this growth is to slice the universe by hypersurfaces of time $t$ measured in units of $H^{-1}$. In this case, volume of all parts of the universe during time $\Delta t = 1$ grows $e^{3}$ times. One can describe this effect by adding the terms $3P_{i}$ to the r.h.s. of Eqs. (\ref{v1}), (\ref{v2}) \cite{Linde:2006nw}. As a result, the functions $P_{i}$  describing the total volume of the  different dS vacua will grow exponentially even in the presence of the sink (if the rate of decay to the sink  is not too large).  In this case, the  functions $P_{i}$ will correspond to the  'pseudo-comoving' probability distribution  \cite{Linde:2006nw}. In this paper we will be interested only in the ratios of the volumes of dS spaces, $P_{i}/P_{j}$. These ratios are not affected by the overall growth of all parts of the universe, and therefore the ratios $P_{i}/P_{j}$ are the same for the comoving and pseudo-comoving probability distributions.  Therefore for simplicity  we will not add  the terms $3P_{i}$ to the r.h.s. of our equations, i.e. we will study comoving probabilities.

To analyze the solutions of equations (\ref{v1}) and (\ref{v2}), let
us first understand the relations between their parameters. Since
entropy of dS space is inversely proportional to energy density, the
entropy of the lower level is highest, ${\bf S_{1}} > {\bf
  S_{2}}$.  As the tunneling is exponentially suppressed,  we have
${\bf S_{2}} > |S(\phi)|$, so we obtain a hierarchy $ {\bf S_{1}} > {\bf
  S_{2}} > |S(\phi)|$, and therefore    $\Gamma_{12}\ll \Gamma_{21}\ll
1$. We will often associate the lower vacuum with our present vacuum
state, where $S_{1} \sim 10^{{120}}$.

For simplicity, we will study here the  possibility that only the
lower vacuum can tunnel to the sink, $\Gamma_{2s}= 0$. The two
equations (\ref{v1}) and (\ref{v2}) can then be solved exactly to
give the general solution
\begin{equation}
P_1 = C_1 e^{-\frac{1}{2}(\Gamma_{12} + \Gamma_{1s} + \Gamma_{21})(t -
  t_0)} \cosh \left[ \frac{A}{2} (t-t_0) \right],
\end{equation}
where $C_1$ and $t_0$ are constants and $A^2 \equiv (\Gamma_{12} +
     \Gamma_{1s} + \Gamma_{21})^2-4 \Gamma_{1s} \Gamma_{21}.$
     Substituting back into the $\dot{P}_1$ equation then gives the ratio
\begin{eqnarray}
\frac{P_2}{P_1} &=& \frac{\Gamma_{12}-\Gamma_{21}+\Gamma_{1s}+A \tanh
  \left[ \frac{A}{2} (t-t_0) \right]}{2 \Gamma_{21}} \nonumber \\
&\rightarrow&  \frac{\Gamma_{12}-\Gamma_{21}+\Gamma_{1s}+A}{2 \Gamma_{21}}
\label{2dsrat}
\end{eqnarray}
as $t \rightarrow \infty$.  These solutions show us that in a comoving
coordinate system, in the
presence of a sink, both of $P_1$ and $P_2$ are decaying, whilst
their ratio approaches a constant value (i.e. their rates of decay
become equal).

One may consider two interesting regimes, providing two very
different types of solution. Suppose first that $\Gamma_{1s} \ll
\Gamma_{21}$, i.e. the probability to fall to the sink from the
lower vacuum is smaller than the probability of the decay of the
upper vacuum.  In this case one recovers the result obtained without
the sink in the previous section:
\begin{equation}
\label{HHeq} {P_{2}\over P_{1}} = {\Gamma_{12}\over \Gamma_{21}} = e^{{\bf S_{2}}-{\bf S_{1}}} \ll 1 .
\end{equation}
It is interesting that this thermal equilibrium is maintained even in
the presence of a sink if $\Gamma_{1s} \ll \Gamma_{21}$. Note that the
required condition for thermal equilibrium is not  $\Gamma_{1s} \ll
\Gamma_{12}$, as one could naively expect, but rather $\Gamma_{1s} \ll
\Gamma_{21}$. We will call such sinks narrow.

Now let us consider the opposite regime, and assume that the decay
rate of the uplifted dS vacuum to the sink is relatively large,
$\Gamma_{1s} \gg \Gamma_{21}$, which automatically means that $\Gamma_{1s}
\gg \Gamma_{12}$.  In this ``wide sink'' regime  the solution of Eq. (\ref{2dsrat}) is
\begin{equation}\label{ooo}
{P_{2}\over P_{1}} =  {\Gamma_{1s}\over \Gamma_{21}} =
e^{{\bf S_{2}}-|S(\phi)| -C_{1}} \approx e^{{\bf S_{2}}-|S(\phi)|} \gg
1 ,
\end{equation}
i.e. one has  {\it an inverted probability distribution}. This
result has a simple interpretation: if the ``thermal exchange''
between the two dS vacua occurs very slowly as compared to the rate
of the decay of the lower dS vacuum, then the main fraction of the
volume of the dS vacua will be in the state with the higher energy
density, because everything that flows to the lower level rapidly
falls to the sink.

\section{Is the Landscape Transversable?} \label{newsec}

Before discussing the details of how AdS sinks can effect thermal
structure in the landscape, we will make a note of another scenario
that can arise in their presence:  The separation of the landscape
into disconnected regions, or islands.

\begin{figure}[ht]
\center \epsfig{file=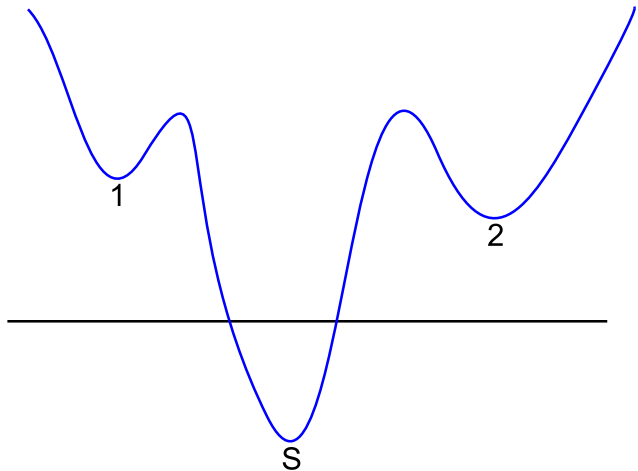,height=6.5cm}
\caption{{\protect {\textit{A potential with two dS minima seperated by a sink.}}}}
\label{2ds1ads2}
\end{figure}

Consider the simple potential shown in Fig. \ref{2ds1ads2}. One may
naively expect that all energetically favorable transitions between
vacua should occur; however, this may not be the case in the
presence of sinks.  In the absence of gravity we know that quantum
mechanical tunneling can only occur from higher energy vacua to
lower energy ones.  In such a picture all vacua slowly tunnel to
lower and lower energy, until they eventually end up in the lowest
energy ground state.  Including the effects of gravity changes this
situation by allowing the possibility of tunneling upwards, to
higher energy.  Although going upwards is less probable than going
down, between any two vacua, the system soon settles down into a
situation where the fluxes going up and down are equal.  This
behavior occurs when the population in the lower energy state far
exceeds the number in the high energy state, compensating the
unlikeliness of jumping upwards by increasing the number of vacua
that could potentially make this transition.  This process is known
as recycling, and, for a system of dS vacua only, allows the whole
system to approach thermal equilibrium, where the ratios of
occupation numbers of any two vacua are given by the equilibrium
ratios (\ref{weinb2}).

AdS and Minkowski sinks, which can be tunneled to, but not back
from, spoil this recycling mechanism and can disrupt the global
thermal structure that would otherwise exist.  One manifestation of
this effect is the altered occupation fractions $P_i/P_j$, discussed
in the previous section; another is the possibility of these sinks
separating the landscape into isolated regions, which may be in
thermal equilibrium internally, but not with each other.  We call such
regions `islands'.

Fig. \ref{2ds1ads2} shows a simple situation which may occur in the
landscape; two dS vacua with an AdS sink between them.  If tunneling
is allowed between the two dS vacua, then thermal equilibrium
between them can occur.  Conversely, if tunneling is not allowed
then these two vacua will be totally isolated from one another.

Consider first the possibility of a Coleman-De Luccia transition from
vacuum 1 to vacuum 2.  As was shown in \cite{Hawking:1981fz}, the CDL
instantons exist only for $V'' \gtrsim V$, where tunneling is
usually assumed to occur from one side of a potential barrier to the
other, as in Fig. \ref{2min}.  The scalar field then rolls down the
potential to the minimum, where reheating may occur if conditions
are right.  For the situation shown in Fig. \ref{2ds1ads2}, however,
such rolling will be down to an AdS minimum, which corresponds to a
collapsing open universe. One may speculate about the
possibility of a tunneling from a collapsing space, or about the
re-emergence of the universe after the collapse, but we do not know
how to study this regime in a controllable way.\footnote{We are grateful to Tom Banks for a discussion of this issue.}  Alternatively, one may
try to find the CDL instantons describing a direct tunneling from
vacuum 1 to vacuum 2, jumping over the AdS minimum.

On the other hand, if we consider $V'' \ll V$, then the CDL instantons do not exist. In this case we still have the Hawking-Moss instanton, and its interpretation in stochastic inflation.  However, according to the stochastic interpretation of the HM tunneling, there can be no tunneling directly from 1 to 2.  Recall that we
view this scenario as representing the scalar field tunneling to the
top of the barrier through a sequence of quantum jumps and
subsequently rolling down to an adjacent vacuum. Clearly this is not
a plausible transition between 1 and 2, since the only adjacent
vacuum is the central AdS sink.

Thus, in general one may encounter situations where it is not possible
to travel from one part of the landscape to another. However, it may
happen that whereas the probability of the transitions discussed above
may be extremely small, they cannot be strictly forbidden; our `no
trespassing' result may be just a consequence of  the approximation
used to study such processes.  Moreover, considering a
multi-dimensional potential with many moduli and fluxes will also add
complexity, as tunneling between vacua 1 and 2 could occur in the
other dimensions.  With an increasing number of dimensions it becomes
increasingly likely that tunneling will be able to occur either directly between 1 and 2, or via some
intermediary vacuum.  Despite these caveats, we find it interesting
that the existence of sinks allows for the possibility of the
landscape being separated into isolated islands.

In the following section we will consider the possibility of
tunneling events directly between two islands, separated by a sink.
Such events could be due to a direct tunneling via the Coleman-De
Luccia instantons in the regime $V'' > V$, or due to a sequence of the
tunneling events though intermediate vacua.  We will show that even if
these events are allowed and the islands are not totally isolated,
they may still be thermally isolated from one another as the sink can
disrupt the equilibrium that would otherwise
exist.  However, our analysis leads us to believe that if the number
of dimensions in the landscape is large enough then the existence of
sinks will not be able to disrupt the tunneling in all of these
dimensions, and thermal contact will therefore be maintained.
Furthermore, the `tortoise and the hare' behavior we find indicates
that if a single high-energy vacuum is allowed to decay to vacua on
each of the islands then this vacuum will act as a bridge, restoring
thermal contact.  For a landscape with very many vacua and
dimensions, it therefore seems implausible that thermally isolated
regions should occur.  We discuss this in more detail below.

\section{Non-Trivial Thermal Structure in the Landscape}\label{island}

In this section we will present our results and analysis of more
general situations than the simple one shown in Fig. \ref{2ds1ads}.
We explore the situations in which thermal equilibrium (which is to
say, detailed balance) between vacua is disturbed.  We hope that
this discussion will be of interest to a wider audience, and can be
understood without having to follow lengthy calculations. The
interested reader can then proceed to the appendix, where the
mathematical minutiae are given.

\subsection{The General Equation}

For a system of $n$ vacua and a single sink we can describe the
evolution of the probability measures, $P$, by:
\begin{equation}
\label{master} \frac{d P_i}{dt} = -\sum_{j \neq i} \Gamma_{i j}
P_i+\sum_{j \neq i} \Gamma_{j i} P_j - \Gamma_{i s} P_i.
\end{equation}
This equation is completely general: Single sinks to which
many vacua can decay with differing amplitudes, and multiple sinks to
which individual vacua can decay (again with differing amplitudes)
amount to the same thing.  The exact form of the general solution to
this set of equations is found in Appendix A.1:
\begin{equation*}
P_i=\sum_j c_{ij} e^{-m_j t} \ ,
\end{equation*}
where the $c_{ij}$ are constants of integration, and the $m_j$ are
constants formed from the transition rates, $\Gamma$.  All $P_i$ have
the same functional form, and their coefficients $c_{ij}$ are related
by factors which are functions of the $\Gamma$s only.

At late times the ratios between different vacua then asymptote to
the constant values:
\begin{equation}
r_{ij}\equiv \lim_{t \rightarrow \infty} \frac{P_i}{P_j},
\end{equation}
which can easily be found, using the solutions for $P_i$. Of course,
finding solutions in a generic landscapes
is not easy, so we will restrict ourselves to a number of simple examples.

In section \ref{tunnsink} we found the solution for the potential
illustrated in Fig. \ref{2ds1ads}, and proceeded to find the
asymptotic limit of the ratio $P_2/P_1$ as $t\rightarrow \infty$. We
now find simple estimates for the asymptotic form of the ratios
$P_i/P_j$ in more generic situations.

\subsection{An Extended 1D Potential}
\begin{figure}[ht]
\center \epsfig{file=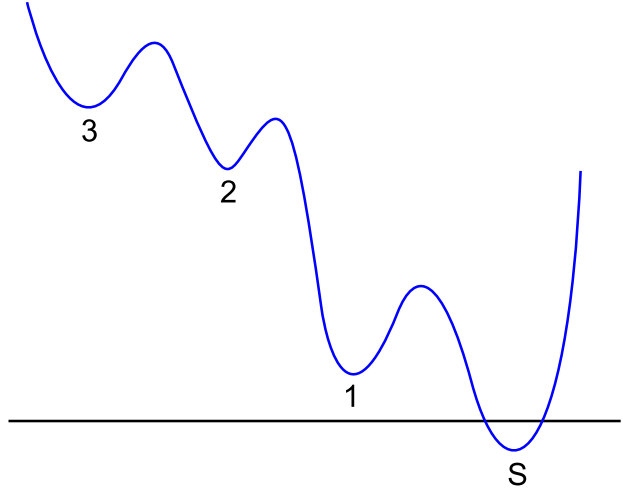,height=6.5cm} \caption{{\protect
{\textit{A one-dimensional potential with three positive and
      one negative minima}}}}
\label{3ds1ads}
\end{figure}

Consider now the extended one-dimensional potential in Fig.
\ref{3ds1ads}.  Here we allow transitions between neighboring de
Sitter minima in both directions, and transitions from 1 to the AdS
sink.  The equations determining the vacuum dynamics, and their
solutions, are given in Appendix A.2.1.

\psfrag{3}{$3$}
For definiteness, we will assume that $V_{3}> V_{2}>V_{1}$, and
therefore $\Gamma_{32}> \Gamma_{23}$, and $\Gamma_{21}>
\Gamma_{12}$. 

Firstly, when $\Gamma_{1s} \ll \Gamma_{21}$, we find:
\begin{equation*}
\frac{P_1}{P_2} \simeq \frac{\Gamma_{21}}{\Gamma_{12}} \gg 1 \qquad \text{and}
\qquad \frac{P_3}{P_2} \simeq \frac{\Gamma_{23}}{\Gamma_{32}} \ll 1.
\end{equation*}
In this limit the
thermodynamic ratios (\ref{weinb2}) are maintained because
$\Gamma_{21} \gg \Gamma_{12}$ and  $\Gamma_{32} \gg \Gamma_{23}$.
When $\Gamma_{1s} \gg \Gamma_{21}$ and
$\Gamma_{21} \ll \Gamma_{32}$, we find:
\begin{equation*}
\frac{P_1}{P_2} \simeq \frac{\Gamma_{21}}{\Gamma_{1s}} \ll 1 \qquad
\text{and} \qquad
\frac{P_3}{P_2} \simeq
\frac{\Gamma_{23}}{\Gamma_{32}} \ll 1.
\end{equation*}
Here the thermal ratio of $P_1$ to $P_2$ is broken, whilst that of
$P_2$ to $P_3$ is maintained.  Lastly, when $\Gamma_{1s} \gg \Gamma_{21}$ and
$\Gamma_{21} \gg \Gamma_{32}$, we find:
\begin{equation*}
\frac{P_1}{P_2} \simeq \frac{\Gamma_{21}}{\Gamma_{1s}} \ll 1 \qquad
\text{and} \qquad
\frac{P_3}{P_2} \simeq \frac{\Gamma_{21}}{\Gamma_{32}} \gg 1.
\end{equation*}
Now the thermal ratio of $P_1$ to $P_2$, and the ratio of $P_2$ to
$P_3$, is broken.

Whilst the equations with three dS vacua are more complicated than
the case for two, it now appears that the interpretation can be
straightforwardly extended.  The ratio $P_2/P_1$ is not affected (to
leading order) by the presence of the extra minimum, and again its
form is prescribed by the magnitude of $\Gamma_{1s}$ relative to
$\Gamma_{21}$.  The new ratio $P_3/P_2$ is found
to take its thermal value if $P_2/P_1$ has a thermal ratio.  If
$P_2/P_1$ has its thermal ratio broken then the rate $\Gamma_{12}$
becomes negligible, and the minimum 1 acts as a sink for the
remaining two dS vacua.  The ratio $P_3/P_2$ can then be calculated
as if they were a system of two dS spaces with vacuum 2 being
allowed to decay to a sink.

\subsection{A Simple ``Multidimensional'' Landscape}\label{multi}

\begin{figure}[ht]
\center \epsfig{file=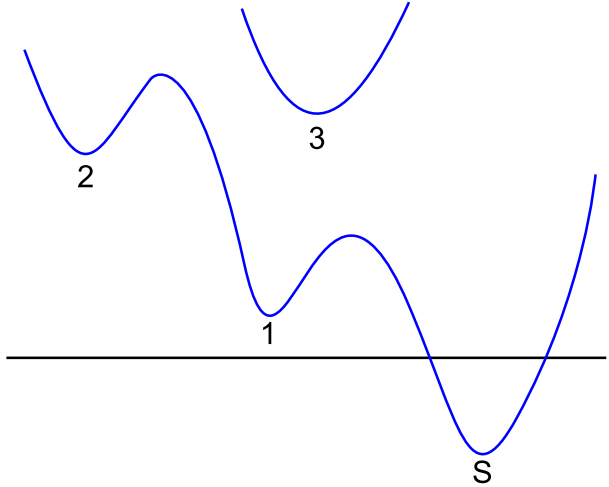,height=6.5cm} \caption{{\protect
{\textit{A multi-dimensional potential with three positive and one
negative minima}}}} \label{3ds1ads2}
\end{figure}

We now wish to consider the case of a multi-dimensional potential,
which we expect to be a slightly more realistic model of the
landscape. Again, we work with an extension of the simple case of
two dS minima and one sink that was given in section \ref{tunnsink}.
Now we consider the potential shown in Fig. \ref{3ds1ads2}.  Here
transitions are allowed between the minima labeled 2 and 3 and the
lower minima 1, which is allowed to decay directly to the sink. This
setup models a simple potential with more than one dimension, i.e.
with more than just pairwise connections between vacua.  The
equations for this potential are given, and solved, in Appendix
A.2.2.

Under the reasonable assumptions that $\Gamma_{21}\gg\Gamma_{12}$ and
$\Gamma_{31}\gg\Gamma_{13}$, we find:
\begin{equation*}
\frac{P_2}{P_1} \simeq
\frac{\Gamma_{12}}{\Gamma_{21}} \ll 1 \qquad \text{and} \qquad \frac{P_3}{P_1}
\simeq \frac{\Gamma_{13}}{\Gamma_{31}} \ll 1
\end{equation*}
when $\Gamma_{21}$ and $\Gamma_{31} \gg \Gamma_{1s}$; whilst for
$\Gamma_{1s}$ and $\Gamma_{31} \gg \Gamma_{21}$ we find:
\begin{equation*}
\frac{P_2}{P_1} \simeq
\frac{\Gamma_{1s}}{\Gamma_{21}} \gg 1 \qquad \text{and} \qquad \frac{P_3}{P_1} \simeq
\frac{\Gamma_{13}}{\Gamma_{31}} \ll 1.
\end{equation*}
The case $\Gamma_{1s}$ and $\Gamma_{21} \gg \Gamma_{31}$ can be
found by symmetry from the above expressions, under the
transcription 2$\leftrightarrow$3.

These results have a straightforward, but slightly
counter-intuitive, interpretation.  When the sink is narrow with
respect to the transition rate $\Gamma_{21}$, as
well as the rate $\Gamma_{31}$, the thermal ratios of both $P_2$ and $P_3$, with respect to
$P_1$, are maintained, as expected.  When the sink is wide with
respect to one of the $\Gamma$s (and narrow with respect to the
other), one of $P_2$ and $P_3$ will maintain its thermal ratio
whilst the other does not, again, as expected.  One could expect that
when the sink is wide with respect to all other transition rates, all
dS vacua will be out of equilibrium. Surprisingly, we have found that
even in this case only one of the upper vacua is out
of its thermal ratio with $P_1$, and the other is not. Furthermore,
the vacuum that is taken out of thermal equilibrium with $P_1$ is
the one with the \textit{slowest} rate of decay.

We interpret this in the following way: If the sink is wide with
respect to all other transition rates, then the magnitude of $P$ in
the vacuum that decays the quickest will quickly become very small.
The slowest decaying vacuum, despite the fact that the rate $\Gamma$
is smaller, will then be the source of the majority of the
probability flux at late-times.  The small flux that is required to
keep the faster decaying vacuum in its thermal ratio with the lower
vacuum is then maintained by the flux from the slower decaying
vacuum, which is forced out of its thermal ratio by the sink.

The picture of the slowest decaying vacuum dominating the
behavior of the universe may seem counter intuitive, but in fact it
has a very simple interpretation: Those who want to survive in a
desert (i.e. near wide sinks) should save water.

Numerical simulations show that this behavior extends to $i>2$ vacua
decaying into a single lower vacuum, which decays to a sink. If the
sink is narrow with respect to all of the other transition rates,
then the asymptotic ratios $P_i/P_1$ approach the thermal ratios
that they would have in the absence of the sink.  If the sink is
wide with respect to any or all of the transition rates to the
higher vacua, then it is only the slowest decaying vacuum that is
forced out of its thermal equilibrium with the lower vacuum.  In this
case, the out-of-equilibrium slowest decaying vacuum will feed all
other vacua, which will be in a state of thermal equilibrium with each
other.

\subsection{Multiple Sinks}

We now have some ideas on how to extend the simple model of two dS
spaces and one AdS space to more general models with multiple
dimensions and higher vacua.  However, so far we have only studied
potentials in which one vacuum is allowed to decay to AdS space.  In
a realistic model of the landscape it is likely that many vacua will
be allowed to decay in this way, and it is the effects of this which
we now study.

\subsubsection{2 dS and 2 AdS}

The potential with two dS minima, shown in Fig. \ref{2ds1ads}, can
be simply extended to allow both of the dS minima to decay to sinks.
This situation is solved in Appendix A.2.3.  Again, we assume
$\Gamma_{12} \ll \Gamma_{21}$.

It can now be shown that if both the sinks are narrow ($\Gamma_{js}
\ll \Gamma_{ij}$) then the ratio $P_2/P_1$ maintains its thermal
value, to leading order.  Similarly, if one sink is narrow, and the
other is wide ($\Gamma_{js} \gg \Gamma_{ij}$), then the leading
order terms are the same as in the absence of the narrow sink.  Now,
if both sinks are wide then the ratio $P_2/P_1$ is determined by the
wider sink.  For example, if we have the hierarchy $\Gamma_{1s} \gg
\Gamma_{2s} \gg \Gamma_{21}$, then
\begin{equation*}
\frac{P_2}{P_1} \simeq \frac{\Gamma_{1s}}{\Gamma_{21}}\ ,
\end{equation*}
whilst for  $\Gamma_{2s} \gg \Gamma_{1s} \gg \Gamma_{21}$, we have
\begin{equation*}
\frac{P_2}{P_1} \simeq \frac{\Gamma_{12}}{\Gamma_{2s}}\ .
\end{equation*}

\subsubsection{3 dS and 2 AdS}\label{sec3ds2ads}

\begin{figure}[ht]
\center \epsfig{file=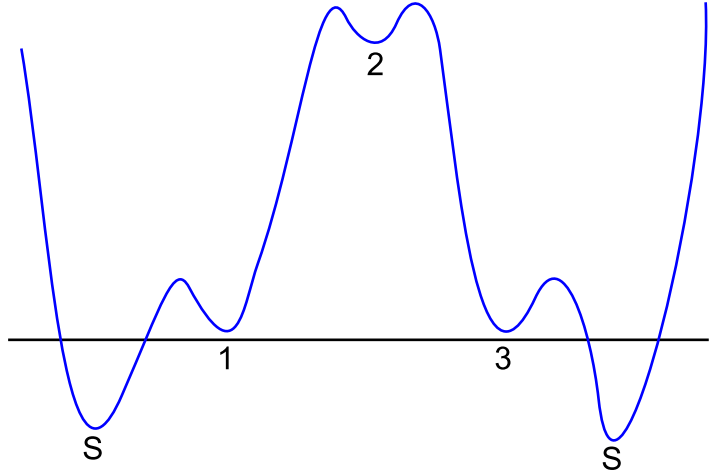,height=6.5cm} \caption{{\protect
{\textit{A potential with three positive minima,
      the lower two of which can decay to sinks.}}}}
\label{3ds2ads}
\end{figure}

We may also consider the case of a single higher vacuum that can
decay to multiple lower vacua, which are, in turn, able to
subsequently decay to sinks. We will model this situation by
considering a potential such as that shown in Fig. \ref{3ds2ads},
where the vacua labeled 1 and 3 are allowed to decay to sinks.  The
equations for this potential are investigated in Appendix A.2.4.  We
now have the heirarchy $\Gamma_{21} \gg \Gamma_{12}$ and $\Gamma_{23}
\gg \Gamma_{32}$.

When both sinks are narrow, $\Gamma_{1s} \ll \Gamma_{21}$ and
$\Gamma_{3s} \ll \Gamma_{23}$, the thermal ratios
\begin{equation*}
\frac{P_1}{P_2} = \frac{\Gamma_{21}}{\Gamma_{12}} \gg 1 \qquad
\text{and} \qquad \frac{P_3}{P_2} = \frac{\Gamma_{23}}{\Gamma_{32}}
\gg 1 \qquad
\end{equation*}
are maintained.  For one narrow sink and one wide ($\Gamma_{is} \gg
\Gamma_{2i}$) the narrow sink becomes irrelevant, and the problem
reduces to the extended one-dimensional potential considered above.

For two wide sinks, if we take
$\Gamma_{1s}\gg\Gamma_{3s}$, without loss of generality, then we
always have the ratio
\begin{equation*}
\frac{P_1}{P_2}=\frac{\Gamma_{21}}{\Gamma_{1s}} \ll 1.
\end{equation*}
The second ratio, $P_3/P_2$, is then determined by the relative magnitudes of
$\Gamma_{3s}$ and $\Gamma_{21}$. If $\Gamma_{3s}\gg\Gamma_{21}$,
then we have
\begin{equation*}
\frac{P_3}{P_2}=\frac{\Gamma_{23}}{\Gamma_{3s}} \ll 1;
\end{equation*}
whilst for $\Gamma_{3s} \ll \Gamma_{21}$ we have
\begin{equation*}
\frac{P_3}{P_2}=\frac{\Gamma_{21}}{\Gamma_{32}} \gg 1.
\end{equation*}
This can be understood in terms
of the previous example of two dS spaces, both connected to sinks.
If the probability charge in vacuum 1 is rapidly depleted by the
fast decay rate to the sink, $\Gamma_{1s}$, then vacuum 1 will
subsequently act as a sink for the remaining two vacua.  If decays
to this new effective sink, $\Gamma_{21}$, are faster than the decay
rate $\Gamma_{3s}$, then this new sink will dominate.  Otherwise,
$\Gamma_{3s}$ will dominate.

\subsubsection{Islands}\label{islands7}

\begin{figure}[ht]
\center \epsfig{file=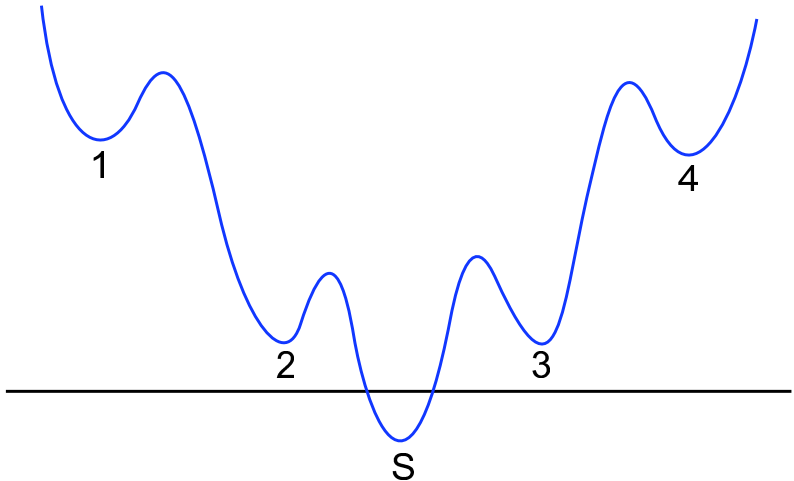,height=6.5cm} \caption{{\protect
{\textit{A potential with two positive minima
      that are allowed to decay to anti-de Sitter space.}}}}
\label{islands}
\end{figure}

Consider the potential shown in Fig. \ref{islands}. Here there are
two dS minima which are allowed to decay to a sink.  (Each of the
lower dS vacua could be considered to decay to different sinks, or
to a common sink, the results are the same).  We have also included
two higher vacua connected to the lower ones, and transitions are
allowed between the two lower minima.  Here we hope to see the
effects of having more than one vacuum decaying to a sink.  We
expect that the presence of the second sink will complicate matters
and allow for the possibility of two `islands' that are in thermal
equilibrium internally, but not with each other.

\psfrag{4}{$4$}

With our knowledge of the general solution, we know that the
asymptotic attractor solutions will be of the form:
\begin{eqnarray*}
P_2 &=& c_1 e^{-m t} \, , \\
P_3 &=& n c_1 e^{-m t} = n P_2\ .
\end{eqnarray*}
As shown below, under the reasonable assumption that $\Gamma_{34}
\ll \Gamma_{43}$ and $\Gamma_{21} \ll \Gamma_{12}$, we find three
possible values of $n$ and $m$ in Appendix A.2.5. In order to choose which solution
for $n=P_3/P_2$ is the relevant one, for a given set of $\Gamma$s,
we must evaluate which $m$ is the smallest.
This $m$ will correspond to the slowest decaying mode, which
dominates in the limit $t\rightarrow\infty$; the corresponding $n$
will then give the appropriate asymptotic ratio of $P_2/P_3$.  A
flow chart is given in Appendix \ref{flowchart}, which can be used
to quickly identify the relevant asymptotic form of $n$ for any
given set of $\Gamma$s.

From the flow chart it can be seen that a necessary condition to
maintain a thermal ratio between vacua 2 and 3 is that either
$\Gamma_{32} \gg \Gamma_{3s}$ or $\Gamma_{23} \gg \Gamma_{2s}$.  A
sufficient condition for maintaining this ratio is either
$\Gamma_{23}$ or $\Gamma_{32} \gg \Gamma_{2s}$ and $\Gamma_{3s}$.
Conversely, a necessary condition for breaking the thermal ratio
between 2 and 3 is either $\Gamma_{2s}$ or $\Gamma_{3s} \gg
\Gamma_{23}$ and $\Gamma_{32}$; whilst a sufficient condition is
given by $\Gamma_{2s}$ and $\Gamma_{3s} \gg \Gamma_{23}$ and
$\Gamma_{32}$.

Two sets of solutions are relatively simple:
\begin{eqnarray*}
n_1 &=& \frac{\Gamma_{23}}{\Gamma_{32} +\Gamma_{3s}} \ , \\
n_2 &=& \frac{\Gamma_{23} +\Gamma_{2s}}{\Gamma_{32}}\ ,
\end{eqnarray*}
with the corresponding values of $m$ being given by
\begin{eqnarray*}
m_1 &=& \Gamma_{12} \ , \\
m_2 &=& \Gamma_{43}\ .
\end{eqnarray*}
The expressions for $m_3$ and $n_3$ are somewhat more complicated,
but they can be simplified depending on the relative
magnitudes of $\Gamma_{23}$, $\Gamma_{32}$, $\Gamma_{2s}$ and
$\Gamma_{3s}$. If $\Gamma_{23}$ or $\Gamma_{32}$ are the fastest of
these rates then the leading order contributions to $m_3$ and $n_3$
are
\begin{equation*}
n_{3a} \simeq \frac{\Gamma_{23}}{\Gamma_{32}} \qquad \text{and}
\qquad m_{3a} \simeq  \frac{\Gamma_{2s} \Gamma_{32} +\Gamma_{3s}
\Gamma_{23}}{
  \Gamma_{23} +\Gamma_{32}},
\end{equation*}
if $\Gamma_{2s}$ is the fastest then
\begin{equation*}
n_{3b} \simeq \frac{\Gamma_{2s}}{\Gamma_{32}} \qquad \text{and}
\qquad m_{3b} \simeq \Gamma_{32} + \Gamma_{3s}
\end{equation*}
and, similarly, if $\Gamma_{3s}$ is the fastest then
\begin{equation*}
n_{3c} \simeq \frac{\Gamma_{23}}{\Gamma_{3s}} \qquad \text{and}
\qquad m_{3c} \simeq \Gamma_{23} + \Gamma_{2s}\ .
\end{equation*}

We will now consider the significance of each of these three
solutions. We take solutions $n_1$ and $m_1$ as corresponding to the
situation in which the probability flux is sourced by the slowly
decaying vacuum 1; solutions $n_2$ and $m_2$ correspond to the
situation in which vacuum 4 is the slowest decaying, and therefore
sources the probability flux; and solutions $n_3$ and $m_3$
correspond to the situation where the majority of the probability
flux is from the lower vacua, 2 and 3.  This interpretation is
supported by the various values of $m$.  For $m_1$ and $m_2$ it can
be seen immediately that the asymptotic rate of decay of the $P_i$
is prescribed by the rate of decay from 1 to 2 or from 4 to 3,
respectively.  This strongly suggests that the vacuum dynamics of
the systems corresponding to these solutions are dominated by the
flux out of vacua 1 and 4.

The interpretation of the $m_3$/$n_3$ solution is a little less
straightforward due to its more complicated form. When the rate of
decay $\Gamma_{2s}$ is fast, $m_{3b}$ is the relevant solution. It
can be seen that $m_{3b}$ is determined by the probability flux out
of vacuum 3 (recall that $\Gamma_{34}$ has been neglected).  By
symmetry, when $\Gamma_{3s}$ is large $m_{3c}$ is the relevant
solution, which is given by the rate of decay out of vacuum 2. When
both rates of decay to the sink are small, $m_{3a}$ is the relevant
solution, which can be seen to be some weighted flux out of both of
the vacua 2 and 3 (analogous to the `reduced mass' of two body
dynamics).  This justifies the above statement that the solutions
$m_3$ and $n_3$ correspond to a system which is dominated by the
flux out of vacua 2 and/or 3.

We again see that the asymptotic form of the `vacuum dynamics' is
not prescribed by the fastest decaying vacua, as may have been
naively expected, but by the slowest.  These vacua hold the majority
of the comoving volume
at late times, and so dominate the late
time evolution of the universe.

The late-time evolution of $P_2$ and $P_3$ straightforwardly gives
the behavior of $P_1$ and $P_4$. Using the notation $p \equiv
P_1/P_2$ and $r \equiv P_4/P_3$, we obtain
\begin{equation*}
p_1 \neq \frac{\Gamma_{21}}{\Gamma_{12}}, \qquad p_2 =
\frac{\Gamma_{21}}{\Gamma_{12}} \qquad \text{and} \qquad p_3 =
\frac{\Gamma_{21}}{\Gamma_{12}}
\end{equation*}
and
\begin{equation*}
r_1 = \frac{\Gamma_{34}}{\Gamma_{43}}, \qquad r_2 \neq
\frac{\Gamma_{34}}{\Gamma_{43}} \qquad \text{and} \qquad r_3 =
\frac{\Gamma_{34}}{\Gamma_{43}},
\end{equation*}
where subscripts $i$ denote that the solution corresponds to $m_i$
and $n_i$.  It can be seen that if the probability charge is being
held in vacua 1 or 4, then that vacuum is out of its thermal ratio
with the lower vacuum to which it can decay, and that the two vacua
on the other side of the sink are in their thermal ratio.  If the
probability charge is being held in either or both of the lower
vacua, then both of the higher vacua have thermal ratios with their
respective lower vacua.

Since we have here a large variety of possibilities, let us single out some of the most interesting regimes.  Suppose first  that  $\Gamma_{23} =\Gamma_{32} =0$, as discussed in Section \ref{newsec}. In this case, in the  wide sink regime, vacua 1 and 2 will be out of thermal equilibrium with each other,  vacua 3 and 4 will be out of thermal equilibrium with each other, and branches (1,2) and (3,4) will be totally disconnected.

Now let us establish some contact between these branches. If
  $\Gamma_{23}$ and $\Gamma_{32}$ are sufficiently small, we will have
  branches (1,2) and (3,4) out of thermal equilibrium with each
  other. If vacuum 1 has the slowest decay rate among all vacua, it
  will be out of equilibrium with vacuum 2. However, in this case
  vacuum 4 will be in thermal equilibrium with  vacuum 3, even if
  $\Gamma_{23} $ and $\Gamma_{32} $ are extremely small. Note that the
  transition to vanishing $\Gamma_{23} $ and $\Gamma_{32} $ is
  discontinuous. This paradoxical situation is similar to the one
  encountered earlier: The slowest decaying vacuum dominates the
  evolution of other domains, but this vacuum becomes irrelevant if it
  totally decouples from other vacua. Other possibilities can be read
  from the flow chart in Appendix  \ref{flowchart}.

The model discussed in this section shows explicitly that tunneling to sinks can break the detailed balance between vacua that would otherwise keep different
regions of the landscape in thermal contact.  This provides a
mechanism by which the landscape can be separated into different
thermally isolated parts, each with a different `temperature'.
In a landscape with high enough dimensionality there may be more than one decay channel between 2 and 3. This will make it more likely that thermal contact will be maintained between various parts of the landscape.

\section{Discussion}\label{disc}

We have, in the course of this work, carried out a detailed
exploration of some toy models of the landscape, and studied their thermal properties. Whilst these models are somewhat limited in their scope, they have revealed to us a rich, and sometimes counter-intuitive structure.

In general, it appears that a necessary condition for the disruption
of thermal ratios between vacua is a rapid rate of decay to an AdS
space -- a wide sink. Intriguingly, however, once this condition is
met, the late-time behavior of the system is controlled by the
\emph{slowest} of the tunneling processes.  For example, when we
considered a simplified version of a multi-dimensional landscape
with a single sink (Section \ref{multi}) we found that only the
slowest decaying vacuum was shifted from its thermal ratio with the
lower vacuum:  All other ratios remained in thermal equilibrium.  We
interpret this as being due to the more rapidly decaying vacua
quickly depleting their populations, leaving the slowest decaying
vacuum to be the primary source of probability flux. The late-time
thermal behavior of the multiverse is then determined by the weakest
transitions -- transitions that would otherwise be irrelevant in the
absence of a sink, or if we took the limit of their rate going to
zero.

We see a similarly interesting interplay in our ``islands'' example in Section \ref{islands7}.
Here, once thermal equilibrium is disrupted by a wide sink, the
late-time behavior of the system is again determined by the slowest
decaying vacuum.  In this example the sink can break the thermal
contact between different regions, leaving several thermally
disconnected `islands'.

In a more complex model there may exist intermediate vacua that can act as bridges, restoring thermal
contact between islands.  Such vacua would be required to be able to
decay to both of the islands, with a non-negligible rate.  As we
have shown, quickly decaying vacua are likely to be in thermal
equilibrium with the lower vacua to which they are allowed to decay.
It may happen that for a realistic landscape with many dimensions and very many vacua, the thermally isolated regions will be relatively rare. However, in order to verify this conjecture one would need to  study simultaneously a disruptive effect of a very large number of sinks and a restoring effect of a very large number of bridges.

We anticipate that further exploration along the lines of this paper
may reveal still more interesting features, allowing us a more
complete and better understood picture of the landscape.
\\\\
\leftline{\bf Acknowledgments}

We are grateful to T. Banks, R. Bousso, R. Easther, B. Freivogel,
S. Shenker, L. Susskind and A. Vilenkin for valuable discussions. This work is supported by NSF grant
PHY-0244728. TC is supported by a Lindemann fellowship.

\appendix
\section{Solving the Equations}
In this appendix we will elaborate on some mathematical details of
the results given in the main body of the text.

\subsection{General Solution}
As discussed in section \ref{island}, the equations governing the
evolution of some set of $P$s are:
\begin{equation*}
\frac{d P_i}{dt} = -\sum_{j \neq i} \Gamma_{i j}
P_i+\sum_{j \neq i} \Gamma_{j i} P_j - \Gamma_{i s} P_i.
\end{equation*}
The first term on the right hand side gives the flux out of the
state denoted by $P_i$ to all the other $P_j$, the second gives the
flux into $P_i$ and the third gives the flux out of $P_i$ and into
the sink.  For $n$ different $P$s, we have a set of $n$ coupled
first order ordinary differential equations.  Such a set of
equations can be manipulated into a single linear $n$th order
equation, for any one particular $P_i$, of the form:
\begin{equation}
\sum_{j}^n a_j \frac{d^j P_i}{dt^j}=0\ .
\end{equation}
The coefficients $a_j$ can be written in terms of the transition
rates, $\Gamma$.  This equation has the general solution
\begin{equation}
\label{solution} P_i = \sum_j^n c_j e^{-m_j t}\ ,
\end{equation}
where the $c_j$ are constants of integration and the constants $m_j$
are the $n$ roots of the $n$th order polynomial $\sum_{j}^n a_j
m^j=0$.  All other $P_i$ can then be seen to have the same
functional form, by substitution back into the original set of
equations (\ref{master}).  The coefficients of the different modes
of these other $P_i$ will be completely determined in terms of $c_j$
and the $\Gamma$s.  This gives the general solution to
(\ref{master}), for all $P_i$, with $n$ arbitrary constants.

At late times the dominant mode in the solution (\ref{solution})
will be the one with the smallest exponent, $m$.  As all the $P$s
have the same functional forms, the same mode will dominate the
evolution of each $P$ at late times, and so the ratios $P_i/P_j$
will asymptotically approach constant values.  The form of the
dominant mode can now be calculated in terms of the constant ratios
$q_{ij} \equiv P_i/P_j$ by adding all the equations (\ref{master})
to get
\begin{equation}
\Big( 1+ \sum_i q_{ij}\Big) \dot{P}_j = -\Big( \Gamma_{j s}
+\sum_i \Gamma_{i s} q_{ij} \Big) P_j\ ,
\end{equation}
which can be integrated to
\begin{equation}
\ln P_j = -\frac{(\Gamma_{j s} +\sum_i \Gamma_{i s} q_{ij})}{(1+ \sum_i q_{ij})}
(t-t_0)\ ,
\end{equation}
where $t_0$ is an arbitrary constant.  We now see that the smallest
value of $m$, which corresponds to the dominant term as
$t\rightarrow \infty$, is given by
\begin{equation}
m =  \frac{(\Gamma_{j s} +\sum_i \Gamma_{i s} q_{ij})}{(1+ \sum_i q_{ij})} < 0.
\end{equation}
Therefore the real part of all of the roots $m_i$ are $>0$, and all
the modes are decaying.

We now know that the late-time attractor solution for each of the
$P$s is an exponentially decaying function, and that this rate of
decay is the same for each $P$.  This allows us to work out the
constant ratios $q_{ij}$, that are asymptotically approached as $t
\rightarrow \infty$.  Taking the mode with the smallest $m$ and
substituting it back into (\ref{master}) gives us a set of $n-1$
algebraic equations which can be solved for the $n-1$ unknown ratios
$q_{ij}$.  The asymptotic form of the $P$s is now, up to a
normalization, completely solved for.

\subsection{Details of Particular Solutions}

Here we will provide the equations that were used to find the results
in the main body of the text, above.  We will assume that all transition rates, $\Gamma$,
are orders of magnitude different from each other.

\subsubsection{An Extended 1D Potential}
The equations governing the system shown in Fig.
\ref{3ds1ads} are given by
\begin{eqnarray*}
\dot{P}_1 &=& -(\Gamma_{12}+\Gamma_{1s}) P_1 + \Gamma_{21} P_2\\
\dot{P}_2 &=& -(\Gamma_{21}+\Gamma_{23}) P_2 + \Gamma_{12} P_1 +\Gamma_{32} P_3\\
\dot{P}_3 &=& -\Gamma_{32} P_3 +\Gamma_{23} P_2\ .
\end{eqnarray*}
We know that the solutions will be of exponential form, so we
substitute the ansatz
\begin{eqnarray*}
P_2 &= c_1 e^{-m t}\\
P_1 &= p c_1 e^{-m t} &= p P_2\\
P_3 &= q c_1 e^{-m t} &= q P_2\ ,
\end{eqnarray*}
and find the expressions
\begin{eqnarray}
\label{pq1} 0 &=& \Gamma_{23} + q (\Gamma_{23} + \Gamma_{21} -
\Gamma_{32}-p
\Gamma_{12}) - \Gamma_{32} q^2\\
0 &=& \Gamma_{21} + p (\Gamma_{21} - \Gamma_{12} - \Gamma_{1s} +
\Gamma_{23} - q \Gamma_{32}) - \Gamma_{12} p^2\ , \nonumber
\end{eqnarray}
or
\begin{eqnarray}
\label{pq2}
0 &=& (\Gamma_{32} q - \Gamma_{23}) (1 + q+p) - \Gamma_{1s} p q \\
0 &=& (\Gamma_{12} p - \Gamma_{21}) (1 + q+p) + \Gamma_{1s} p (1+q)\
.
 \nonumber
\end{eqnarray}

We find three sets of
solutions. Firstly, when $\Gamma_{1s} \ll$ either $
\Gamma_{12}$ or $\Gamma_{21}$, we see from equations (\ref{pq1}) that:
\begin{equation*}
p \simeq \frac{\Gamma_{21}}{\Gamma_{12}} \gg 1 \qquad \text{and}
\qquad q \simeq \frac{\Gamma_{23}}{\Gamma_{32}},
\end{equation*}
where we have taken $\Gamma_{21} \gg \Gamma_{12}$. In this limit the
thermodynamic ratios (\ref{weinb2}) are maintained. For $p\ll1$ and
$q\ll1$, $p$ and $q$ are given by (\ref{pq2}) as:
\begin{eqnarray*}
p &\simeq& \frac{\Gamma_{21}}{\Gamma_{12}+\Gamma_{1s}} \simeq \frac{\Gamma_{21}}{\Gamma_{1s}} \ll 1\\
q &\simeq& \frac{(\Gamma_{12} + \Gamma_{1s})
\Gamma_{23}}{\Gamma_{12}\Gamma_{32} + \Gamma_{1s}
(\Gamma_{32}-\Gamma_{21})} \simeq
\frac{\Gamma_{23}}{\Gamma_{32}-\Gamma_{21}} \simeq
\frac{\Gamma_{23}}{\Gamma_{32}} \ll 1
\end{eqnarray*}
where ($\Gamma_{1s} \gg \Gamma_{12}$ and $\Gamma_{21}$) and
($\Gamma_{32} \gg \Gamma_{21}$ and $\Gamma_{23}$). Here the thermal
ratio of $P_1$ to $P_2$ is broken, whilst that of $P_2$ to $P_3$ is
maintained.  Lastly, for $p\ll 1$ and $q \gg 1$, the solutions to
(\ref{pq2}) are:
\begin{eqnarray*}
p &\simeq& \frac{\Gamma_{21}}{\Gamma_{12}+\Gamma_{1s}} \simeq  \frac{\Gamma_{21}}{\Gamma_{1s}} \ll 1\\
q &\simeq& \frac{\Gamma_{12} \Gamma_{23} + \Gamma_{1s} (\Gamma_{21}
+
  \Gamma_{23})}{(\Gamma_{12} + \Gamma_{1s}) \Gamma_{32}} \simeq
\frac{\Gamma_{21}+\Gamma_{23}}{\Gamma_{32}} \gg 1
\end{eqnarray*}
where ($\Gamma_{1s} \gg \Gamma_{12}$ and $\Gamma_{21}$) and
($\Gamma_{32} \ll$ either $ \Gamma_{21}$ or $\Gamma_{23}$). Again, the thermal
ratio of $P_1$ to $P_2$ is broken and now the ratio of $P_2$ to
$P_3$ is thermal if $\Gamma_{21} \ll \Gamma_{23}$ and not thermal if
$\Gamma_{21} \gg \Gamma_{23}$.  Note, that no solutions were found
for $p$ and $q \gg 1$, when the sink is wide.  These results have been
verified numerically.

\subsubsection{A Simple Multi-Dimensional Potential}
For the setup show in Fig. \ref{3ds1ads2} the system of equations
governing the vacuum dynamics is
\begin{eqnarray*}
\dot{P}_1 &=& -(\Gamma_{12} +\Gamma_{13}+\Gamma_{1s})
P_1+\Gamma_{21}
P_2 + \Gamma_{31} P_3\\
\dot{P}_2 &=& - \Gamma_{21} P_2 +\Gamma_{12} P_1\\
\dot{P}_3 &=& - \Gamma_{31} P_3 +\Gamma_{13} P_1\ .
\end{eqnarray*}
Substituting the ansatz
\begin{eqnarray*}
P_1 &= c_1 e^{-m t}\\
P_2 &= p c_1 e^{-m t} &= p P_1\\
P_3 &= q c_1 e^{-m t} &= q P_1\ ,
\end{eqnarray*}
gives the equations
\begin{eqnarray*}
p &=& \frac{(\Gamma_{31} q - \Gamma_{13}) (1 +
q)-q\Gamma_{1s}}{\Gamma_{13} - \Gamma_{31} q}\\
q &=& \frac{(\Gamma_{21} p - \Gamma_{12}) (1 +
p)-p\Gamma_{1s}}{\Gamma_{12} - \Gamma_{21} p}\ .
\end{eqnarray*}
Under the reasonable
assumptions that $\Gamma_{21}\gg\Gamma_{12}$ and
$\Gamma_{31}\gg\Gamma_{13}$, for $p\ll1$ and $q\ll1$ we have
\begin{equation*}
p \simeq \frac{\Gamma_{12}}{\Gamma_{21} - \Gamma_{1s}} \simeq
\frac{\Gamma_{12}}{\Gamma_{21}} \ll 1 \qquad \text{and} \qquad q
\simeq \frac{\Gamma_{13}}{\Gamma_{31} - \Gamma_{1s}} \simeq
\frac{\Gamma_{13}}{\Gamma_{31}} \ll 1
\end{equation*}
when $\Gamma_{21}$ and $\Gamma_{31} \gg \Gamma_{1s}$; whilst for $p
\gg 1 \gg q$ we find:
\begin{eqnarray*}
p &\simeq& \frac{\Gamma_{12} + \Gamma_{1s}}{\Gamma_{21}} \simeq
\frac{\Gamma_{1s}}{\Gamma_{21}} \gg 1 \\ q &\simeq&
\frac{\Gamma_{13} (\Gamma_{12} + \Gamma_{1s})}{\Gamma_{12}
\Gamma_{31}
  + \Gamma_{1s} \Gamma_{31}- \Gamma_{1s} \Gamma_{21}} \simeq \frac{\Gamma_{13}}{\Gamma_{31}} \ll 1
\end{eqnarray*}
when $\Gamma_{1s}$ and $\Gamma_{31} \gg \Gamma_{21}$.  The case $q
\gg 1 \gg p$ can be found by symmetry from the above expressions,
under the transcription 2$\leftrightarrow$3 and $p \leftrightarrow
q$.  We find no consistent solution exists with $p \gg 1$ and $q \gg
1$. These solutions have been numerically verified in the limits indicated above.

\subsubsection{Two de Sitter and Two Sinks}
The dynamical equations for this situation are given by:
\begin{eqnarray*}
\dot{P}_1 &=& -(\Gamma_{1 2}+\Gamma_{1 s}) P_1 + \Gamma_{21} P_2\\
\dot{P}_2 &=& -(\Gamma_{21}+\Gamma_{2s}) P_2 + \Gamma_{12} P_1\ ,
\end{eqnarray*}
which have the general solution:
\begin{eqnarray*}
\frac{P_2}{P_1} &=& \frac{\Gamma_{12}-\Gamma_{21}+\Gamma_{1s}-\Gamma_{2s}+B \tanh
  \left[ \frac{B}{2} (t-t_0) \right]}{2 \Gamma_{21}} \nonumber \\
&\rightarrow&
\frac{\Gamma_{12}-\Gamma_{21}+\Gamma_{1s}-\Gamma_{2s}+B}{2
\Gamma_{21}}\
\end{eqnarray*}
as $t \rightarrow \infty$.  Here $B^2 \equiv
(\Gamma_{12}+\Gamma_{21}+\Gamma_{1s}+\Gamma_{2s})^2 -4 (\Gamma_{12}
\Gamma_{2s} + \Gamma_{1s} \Gamma_{21} + \Gamma_{1s} \Gamma_{2s}))$ and
the results in the text follow from this asymptotic form.

\subsubsection{Three de Sitter and Two Sinks}

The relevant evolution equations are now:
\begin{eqnarray*}
\dot{P}_1 &=& -(\Gamma_{12}+\Gamma_{1s}) P_1 + \Gamma_{21} P_2\\
\dot{P}_2 &=& -(\Gamma_{21}+\Gamma_{23}) P_2 + \Gamma_{12} P_1 +\Gamma_{32} P_3\\
\dot{P}_3 &=& -(\Gamma_{32}+\Gamma_{3s}) P_3 +\Gamma_{23} P_2\
\end{eqnarray*}
which yield
\begin{eqnarray}
\label{sinks}
0 &=& \Gamma_{23} + q (\Gamma_{23} + \Gamma_{21} - \Gamma_{3s} - \Gamma_{32} - p
\Gamma_{12}) - \Gamma_{32} q^2\\
0 &=& \Gamma_{21} + p (\Gamma_{21} - \Gamma_{12} - \Gamma_{1s} +
\Gamma_{23} - q \Gamma_{32}) - \Gamma_{12} p^2\ , \nonumber
\end{eqnarray}
or
\begin{eqnarray}
\label{sinks2}
(1+q) (p \Gamma_{1s}+q\Gamma_{3s}) &=& (1+p+q) (q
  \Gamma_{3s}-p\Gamma_{12}+\Gamma_{21})\\
(1+p) (p \Gamma_{1s}+q\Gamma_{3s}) &=& (1+p+q) (p
  \Gamma_{1s}-q\Gamma_{32}+\Gamma_{23})
\nonumber
\end{eqnarray}
where $p \equiv P_1/P_2$ and $q \equiv P_3/P_2$. When
both sinks are narrow ($\Gamma_{1s} \ll \Gamma_{12}$ and
$\Gamma_{3s} \ll \Gamma_{32}$) it can be seen from (\ref{sinks}) that the thermal ratios
\begin{equation*}
\frac{P_1}{P_2} = \frac{\Gamma_{21}}{\Gamma_{12}} \gg 1 \qquad
\text{and} \qquad \frac{P_3}{P_2} = \frac{\Gamma_{23}}{\Gamma_{32}}
\gg 1 \qquad
\end{equation*}
are maintained.  For one narrow sink and one wide ($\Gamma_{is} \ll
\Gamma_{i2}$) the narrow sink becomes irrelevant, and the problem
reduces to the one considered above.

The case of two wide sinks now remains. From (\ref{sinks2}), for $p\ll1$ and $q\ll1$ we
obtain:
\begin{eqnarray*}
p &\simeq& \frac{\Gamma_{21}}{\Gamma_{12}+\Gamma_{1s}} \simeq
\frac{\Gamma_{21}}{\Gamma_{1s}} \ll 1\\
q &\simeq& \frac{\Gamma_{23}}{\Gamma_{32}+\Gamma_{3s}} \simeq
\frac{\Gamma_{23}}{\Gamma_{3s}} \ll 1\
\end{eqnarray*}
whilst $p\ll1\ll q$ gives:
\begin{eqnarray*}
p &\simeq& \frac{\Gamma_{21}}{\Gamma_{12}+\Gamma_{1s}} \simeq
\frac{\Gamma_{21}}{\Gamma_{1s}} \ll 1\\
q &\simeq& \frac{\Gamma_{23}+\Gamma_{21}-\Gamma_{3s}}{\Gamma_{32}}
\simeq \frac{\Gamma_{21}}{\Gamma_{32}} \gg 1
\end{eqnarray*}
where
$\Gamma_{32}\ll\Gamma_{23}\ll\Gamma_{3s}\ll\Gamma_{21}\ll\Gamma_{1s}$.
The remaining case $p\gg q\gg1$ has no solutions (given two wide
sinks).

\subsubsection{Islands}
The potential shown in Fig. \ref{islands} is governed by the
set of equations:
\begin{eqnarray*}
\dot{P}_1 &=& -\Gamma_{12} P_1+\Gamma_{21} P_2\\
\dot{P}_2 &=& -(\Gamma_{21} + \Gamma_{23}+\Gamma_{2s}) P_2+
\Gamma_{12} P_1+\Gamma_{32} P_3\\
\dot{P}_3 &=& -(\Gamma_{34} + \Gamma_{32}+\Gamma_{3s}) P_3+
\Gamma_{43} P_4+\Gamma_{23} P_2\\
\dot{P}_4 &=& -\Gamma_{43} P_4+\Gamma_{34} P_3\ .
\end{eqnarray*}
These four coupled first-order equations can be recast into a set of
two coupled second-order ordinary differential equations, for the
variables $P_2$ and $P_3$
\begin{eqnarray*}
\ddot{P}_2+(\Gamma_{12}+\Gamma_{21}+\Gamma_{23}+\Gamma_{2s}) \dot{P}_2
+\Gamma_{12} (\Gamma_{23}+\Gamma_{2s}) P_2 &=& \Gamma_{32} \dot{P}_3+
  \Gamma_{12} \Gamma_{32} P_3\\
\ddot{P}_3+(\Gamma_{43}+\Gamma_{34}+\Gamma_{32}+\Gamma_{3s}) \dot{P}_3
+\Gamma_{43} (\Gamma_{32}+\Gamma_{3s}) P_3 &=& \Gamma_{23} \dot{P}_2+
  \Gamma_{43} \Gamma_{23} P_2\ .
\end{eqnarray*}
We know that the asymptotic attractor solutions have the form
\begin{eqnarray*}
P_2 &=& c_1 e^{-m t}\\
P_3 &=& n c_1 e^{-m t} = n P_2\ ,
\end{eqnarray*}
which on substitution into the second-order equations above, give
\begin{eqnarray*}
\Gamma_{12} (\Gamma_{23}+\Gamma_{2s}-n \Gamma_{32}-m)-m (\Gamma_{21}
+\Gamma_{23}+\Gamma_{2s} - n\Gamma_{32}-m) &=& 0\\
n ( (\Gamma_{32}+\Gamma_{3s}) (\Gamma_{43}-m)-m (\Gamma_{34}
+\Gamma_{43}-m))- \Gamma_{23} (\Gamma_{43}-m) &=&0\ .
\end{eqnarray*}
Under the reasonable assumption that $\Gamma_{34} \ll \Gamma_{43}$
and $\Gamma_{21} \ll \Gamma_{12}$ these two equations can be solved
to give the three possible values of $n$ and $m$ discussed above:
\begin{eqnarray*}
n_1 &=& \frac{\Gamma_{23}}{\Gamma_{32} +\Gamma_{3s}-\Gamma_{12}}\\
n_2 &=& \frac{\Gamma_{23} +\Gamma_{2s}-\Gamma_{43}}{\Gamma_{32}}\\
n_3 &=& \frac{\Gamma_{23}-\Gamma_{32}+\Gamma_{2s}-\Gamma_{3s} +
  \sqrt{(\Gamma_{23}-\Gamma_{32}+\Gamma_{2s}-\Gamma_{3s})^2+4
    \Gamma_{23} \Gamma_{32}}}{2 \Gamma_{32}}\ ,
\end{eqnarray*}
with the corresponding values of $m$ being given by
\begin{eqnarray*}
m_1 &=& \Gamma_{12}\\
m_2 &=& \Gamma_{43}\\
m_3 &=&  \frac{1}{2} \left(\Gamma_{23}+\Gamma_{32}+\Gamma_{2s}+\Gamma_{3s} -
  \sqrt{(\Gamma_{23}-\Gamma_{32}+\Gamma_{2s}-\Gamma_{3s})^2+4
    \Gamma_{23} \Gamma_{32}} \right)\ .
\end{eqnarray*}
There is a fourth mathematically permissible value of $m$ and $n$,
however it corresponds to $m<0$ for all $\Gamma_{ij}$ and so we do
not consider it to be of any physical significance (a negative value
of $P$ makes very little sense).  These results have been confirmed numerically.

\newpage

\section{Flow Chart}\label{flowchart}
\begin{figure}[h!t!]
\center \epsfig{file=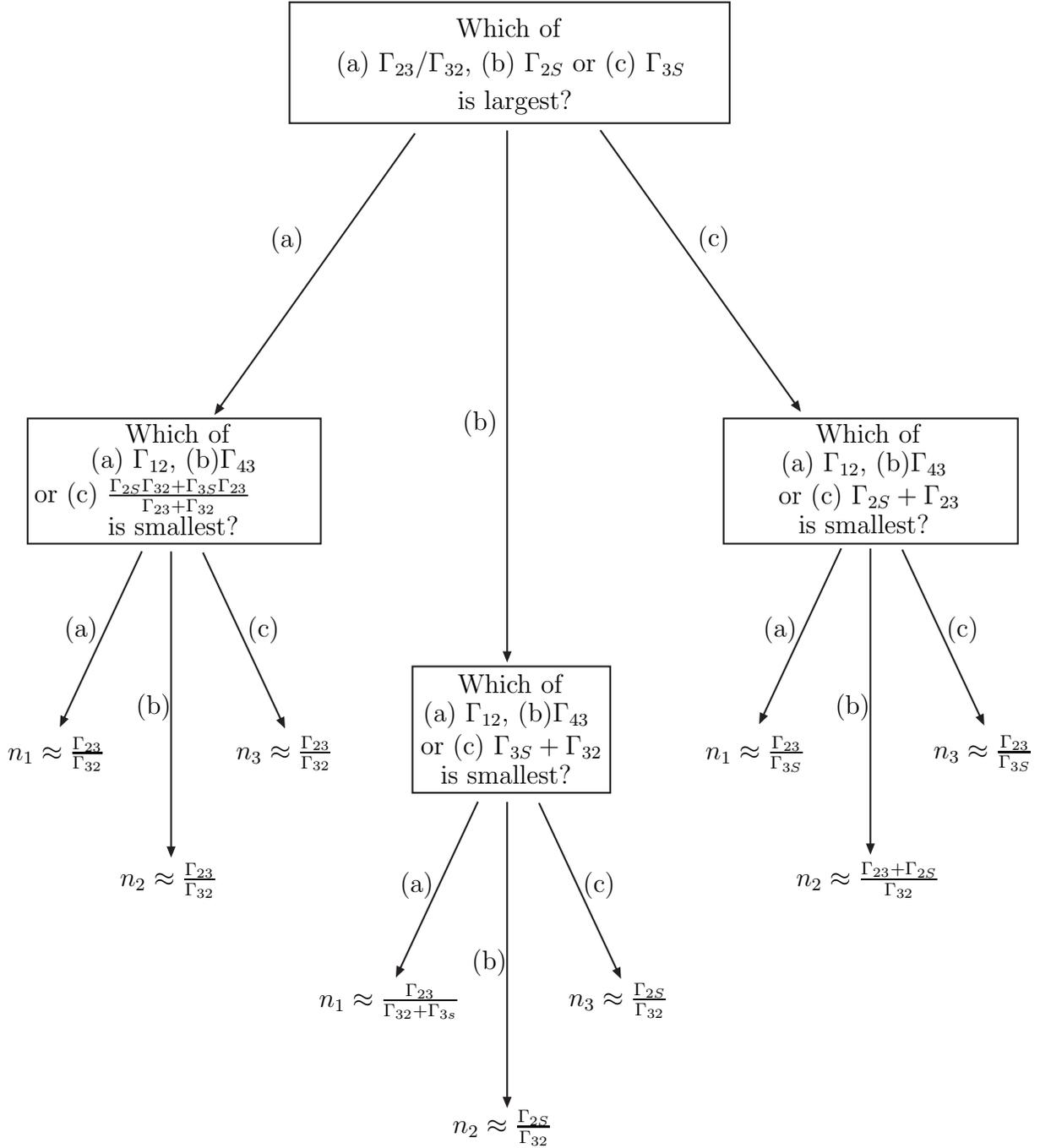,height=18cm}
\caption{{\protect {\textit{Flowchart showing different asymptotic
limits for the ``islands'' example.}}}} \label{flowchart1}
\end{figure}

Fig. \ref{flowchart1} shows the various possible asymptotic limits
for the ``thermal islands'' discussed in section \ref{islands7}.

\psfrag{(a)}{(a)} \psfrag{(b)}{(b)} \psfrag{(c)}{(c)}
\psfrag{W1}{Which of} \psfrag{W2}{(a) $\Gamma_{23}$/$\Gamma_{32}$,
(b) $\Gamma_{2S}$ or (c) $\Gamma_{3S}$} \psfrag{W3}{is largest?}
\psfrag{W21}{Which of} \psfrag{W22}{(a) $\Gamma_{12}$,
(b)$\Gamma_{43}$} \psfrag{W23}{or (c)
$\frac{\Gamma_{2S}\Gamma_{32}+\Gamma_{3S}\Gamma_{23}}{\Gamma_{23}+\Gamma_{32}}$}
\psfrag{W24}{is smallest?} \psfrag{W31}{Which of} \psfrag{W32}{(a)
$\Gamma_{12}$, (b)$\Gamma_{43}$} \psfrag{W33}{or (c)
$\Gamma_{3S}+\Gamma_{32}$} \psfrag{W34}{is smallest?}
\psfrag{W41}{Which of} \psfrag{W42}{(a) $\Gamma_{12}$,
(b)$\Gamma_{43}$} \psfrag{W43}{or (c) $\Gamma_{2S}+\Gamma_{23}$}
\psfrag{W44}{is smallest?}
\psfrag{m1}{$n_1\approx\frac{\Gamma_{23}}{\Gamma_{32}}$}
\psfrag{m12}{$n_2\approx\frac{\Gamma_{23}}{\Gamma_{32}}$}
\psfrag{m13}{$n_3\approx\frac{\Gamma_{23}}{\Gamma_{32}}$}
\psfrag{m21}{$n_1\approx\frac{\Gamma_{23}}{\Gamma_{32}+\Gamma_{3s}}$}
\psfrag{m22}{$n_2\approx\frac{\Gamma_{2S}}{\Gamma_{32}}$}
\psfrag{m23}{$n_3\approx\frac{\Gamma_{2S}}{\Gamma_{32}}$}
\psfrag{m31}{$n_1\approx\frac{\Gamma_{23}}{\Gamma_{3S}}$}
\psfrag{m32}{$n_2\approx\frac{\Gamma_{23}+\Gamma_{2S}}{\Gamma_{32}}$}
\psfrag{m33}{$n_3\approx\frac{\Gamma_{23}}{\Gamma_{3S}}$}

\end{document}